\documentclass[a4paper]{mn2e}
\usepackage{natbib_jrm,times}
\usepackage{graphicx}
\bibpunct[, ]{(}{)}{;}{a}{}{,}

\def \mnras {MNRAS}
\def \pasp {PASP}
\def \apj {ApJ}

\def \apjl {ApJL}
\def \aap {A\&A}

\def \araa {ARAA}
\def \iaucirc {IAUC}

\def \apss {Ap\&SS}

\newcommand{\kms} {$\mathrm{ km \; s^{-1}}\,$}

\def\lesssim{\mathrel{\hbox{\rlap{\hbox{\lower4pt\hbox{$\sim$}}}\hbox{$<$}}}}
\def\gtrsim{\mathrel{\hbox{\rlap{\hbox{\lower4pt\hbox{$\sim$}}}\hbox{$>$}}}}
\newcommand{\halpha} {$\mathrm{H\alpha}$}

\long\def\symbolfootnote[#1]#2{\begingroup%
\def\thefootnote{\fnsymbol{footnote}}\footnote[#1]{#2}\endgroup} 

\begin{document}
\title[SN~2012ec]{Supernova 2012ec: Identification of the progenitor and early monitoring with PESSTO\thanks{Based on observations made with ESO Telescopes at the La Silla Paranal Observatory under programmes 089.D-0305 \& 188.D-3003}}
\author[Maund et al.]{
\parbox[t]{\textwidth}{\raggedright
J.R. ~Maund$^{1,2,3}$\thanks{Email: j.maund@qub.ac.uk}, 
M. Fraser$^{1}$, 
S.~J. Smartt$^{1}$,
M.T. Botticella$^{4}$,
C. Barbarino$^{4}$,
M. Childress$^{5}$,
A. Gal-Yam$^{6}$,
C. Inserra$^{1}$,
G. Pignata$^{7}$,
D. Reichart$^{8}$,
B. Schmidt$^{5}$,
J. Sollerman$^{9}$,
F. Taddia$^{9}$,
L. Tomasella$^{10}$,
S. Valenti$^{11,12}$,
\& O. Yaron$^{6}$
}
\vspace*{6pt}\\
$^{1}$ Astrophysics Research Centre, School of Mathematics and Physics, Queen's University Belfast, Belfast, BT7 1NN, Northern Ireland, U.K.\\
$^{2}$ Dark Cosmology Centre, Niels Bohr Institute, University of Copenhagen, Juliane Maries Vej 30, 2100 Copenhagen, DK.\\
$^{3}$ Royal Society Research Fellow\\
$^{4}$ INAF - Osservatorio astronomico di Capodimonte, Salita Moiariello 16, I- 80131 Napoli, Italy\\
$^{5}$ Research School of Astronomy and Astrophysics, Australian National University, Cotter Road, Weston Creek, ACT 2611, Australia\\
$^{6}$  Department of Particle Physics and Astrophysics, The Weizmann Institute of Science, Rehovot 76100, Israel\\
$^{7}$ Departamento de Ciencias Fisicas, Universidad Andres Bello, Avda. Republica 252, Santiago, Chile\\ 
$^{8}$ Department of Physics and Astronomy, University of North Carolina at Chapel Hill, 120 E. Cameron Ave., Chapel Hill, NC 27599, U.S.A.\\
$^{9}$ The Oskar Klein Centre, Department of Astronomy, AlbaNova,
Stockholm University, 10691 Stockholm, Sweden\\
$^{10}$ INAF - Osservatorio Astronomico di Padova, Vicolo dellOsservatorio 5, 35122, Padova, Italy\\
$^{11}$ Las Cumbres Observatory Global Telescope Network, 6740 Cortona Dr., Suite 102, Goleta, CA 93117, USA\\
$^{12}$ Department of Physics, University of California, Santa Barbara, Broida Hall, Mail Code 9530, Santa Barbara, CA 93106-9530, USA
}
\maketitle
\begin{abstract}
We present the identification of the progenitor of the Type IIP
SN~2012ec in archival pre-explosion {\it HST} {\it WFPC2} and {\it
  ACS/WFC} {\it F814W} images.  The properties of the progenitor are
further constrained by non-detections in pre-explosion {\it WFPC2}
{\it F450W} and {\it F606W} images.  We report a series of early
photometric and spectroscopic observations of SN~2012ec.  The
$r^{\prime}$-band light curve shows a plateau with
$M_{r^{\prime}}=-17.0$.  The early spectrum is similar to the Type
IIP SN~1999em, with the expansion velocity measured at
$\mathrm{H\alpha}$ absorption minimum of
$-11\,700\,\mathrm{km\,s^{-1}}$ (at 1 day post-discovery).  The
photometric and spectroscopic evolution of SN~2012ec shows it to be a Type IIP SN, discovered only a few days post-explosion
($<6\mathrm{d}$).  We derive a luminosity for the progenitor, in
comparison with MARCS model SEDs, of $\log L/L_{\odot} = 5.15\pm0.19$,
from which we infer an initial mass range of $14-22M_{\odot}$.  This is the
first SN with an identified progenitor to be followed by the Public
ESO Spectroscopic Survey of Transient Objects (PESSTO).
\end{abstract}

\section{Introduction}
\label{intro}
The hydrogen-rich Type IIP supernovae (SNe) are the most common type
of SN in the local universe
\citep{2011MNRAS.412.1473L,2001ibsp.conf..199C}.  The standard
prediction of stellar evolution models is that such SNe arise from the
lowest mass stars to end their lives as core-collapse (CC) SNe, whilst
still retaining their massive hydrogen envelopes that give rise to the
characteristic light curve plateau \citep{heg03,eld04}.  Red
supergiants have been directly observed at the sites of these SNe in
fortuitous pre-explosion images \citep[see][for a
  review]{2009ARA&A..47...63S} and, with the observation of the
disappearance of the progenitor stars in late-time post-explosion
images, this scenario for the production of Type IIP SNe has been
confirmed \citep{2009Sci...324..486M}.  A number of key problems are
still outstanding in our understanding of Type IIP SNe, in particular
the diversity in the explosion energies and the amounts of $^{56}$Ni produced in these events \citep{pasto2005cs}, and the apparent deficit of
observed progenitors with $M_{ZAMS} \gtrsim 16M_{\odot}$ (termed the
``Red Supergiant Problem''; \citealt{2008arXiv0809.0403S}).\\ We
report the identification of the progenitor of the Type IIP SN~2012ec
in pre-explosion {\it HST} {\it WFPC2} and {\it ACS}
images, and the early photometric and spectroscopic properties of the
SN. SN~2012ec was discovered by \citet{2012ec:disc} on 2012 Aug
11.039UT.  The SN is located $0.7\arcsec$E and $15.9\arcsec$N of the
centre of the host galaxy NGC~1084.  A spectrum of the SN acquired on
2012 Aug 12 showed it to be a young Type IIP SN, a few days
post-explosion \citep{2012ec:spec}.  NGC~1084 has previously hosted 4
SNe: 2009H \citep{2009CBET.1656....1L}, 1998dl
\citep{1998IAUC.6992....1K}, 1996an \citep{1996IAUC.6442....1N} and
1963P.\\ 
The Tully-Fisher distance \citep{2008ApJ...676..184T}, quoted from the
Extragalactic Distance Database\footnote{http://edd.ifa.hawaii.edu/},
is $\mu=31.19\pm0.13$ mag.  The position of SN~2012ec is coincident
with the region C9 identified by \citet{2007MNRAS.381..511R} in
NGC~1084, for which they derive an oxygen abundance of
$12+\log\left(\frac{O}{H}\right)=8.93\pm0.1$.  Assuming a solar
metallicity corresponding to an oxygen abundance of $8.66\pm0.05$
\citep{2004A&A...417..751A}, we adopt a metallicity of
$\log\left({Z}/{Z_{\odot}}\right)\approx +0.27$ at the position of
the SN.  The degree of foreground Galactic reddening is
$E(B-V)=0.024$ mag \citep{2011ApJ...737..103S}\footnote{quoted from
  NED http://ned.ipac.caltech.edu}.
\section{The Early Photometric and Spectroscopic Characteristics of SN~2012ec}
\label{sec:evo}
\begin{figure*}
\includegraphics[angle=0, width=17.5cm]{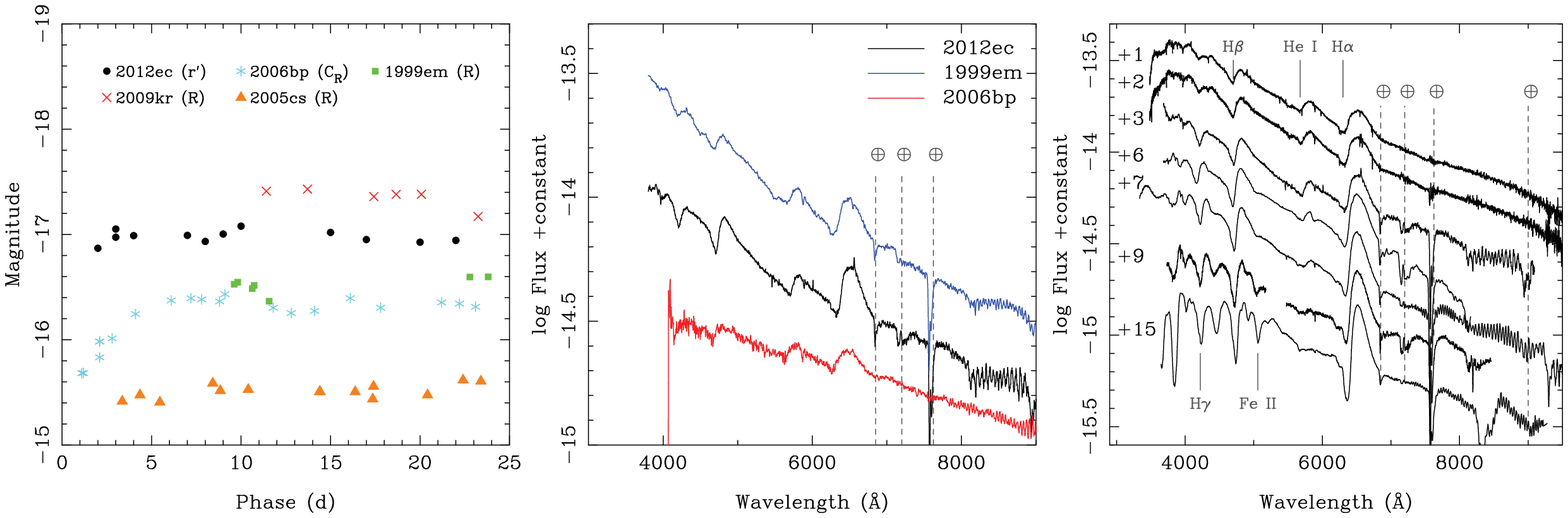}
\caption{The early photometric and spectroscopic characteristics of the Type IIP SN~2012ec. {\it Left Panel)} Early $r^{\prime}$ photometry of SN~2012ec compared to other Type II SNe: 2009kr \citep{2010ApJ...714L.280F}, 2006bp \citep{2007ApJ...666.1093Q}, 2005cs \citep{pasto2005cs}, and 1999em \citep{2001ApJ...558..615H,2003MNRAS.338..939E}.  The phase of the SN~2012ec data is given with respect to the date of discovery.  {\it Centre Panel)} The early spectrum of SN~2012ec compared to the normal Type IIP SNe 1999em \citep{2001ApJ...558..615H,2003MNRAS.338..939E} and 2006bp \citep{2007ApJ...666.1093Q}, at $+7$ days and 9 days post-explosion respectively (the SN spectra have been corrected for the respective recessional velocity of their host galaxies and for foreground Milky Way extinction, as quoted by NED). {\it Right Panel)} The early spectroscopic evolution of SN~2012ec, with the phase of each spectrum given relative to the date of discovery (2012 Aug 11).  Characteristic lines, observed in the spectra Type IIP SNe, are indicated by the solid grey lines, and telluric features are indicated by the dashed  grey lines.}
\label{fig:evo}
\end{figure*}
A photometric and spectroscopic campaign to follow the evolution of
this SN is being conducted by the Public ESO Spectroscopic Survey of
Transient Objects (PESSTO)
collaboration\footnote{http://www.pessto.org}.  Full results of the
observations of this SN will be presented in a future paper; here we
present a subset of these observations to classify and characterise
the early time behaviour of SN~2012ec.
Photometry was collected using the PROMPT telescopes \citep{2005NCimC..28..767R}, and the Liverpool Telescope using RATcam and the Infrared Optical (IO) camera.  The
photometric observations were conducted using Sloan filters, and were
reduced in the standard fashion using {\sc iraf}\footnote{IRAF is
  distributed by the National Optical Astronomy Observatory, which is
  operated by the Association of Universities for Research in
  Astronomy (AURA) under cooperative agreement with the National
  Science Foundation.} and calibrated against stars in the Sloan DR6
catalogue\footnote{http://www.sdss.org/dr6/}.  The $r^{\prime}$-band
light curve (in the Vega magnitude system) is presented in Figure
\ref{fig:evo}, and compared with light curves of other Type II SNe in
similar, but {\it not identical}, bands.  The light curve of SN~2012ec clearly shows a plateau, implying that the SN is a Type IIP SN. The absolute magnitude (corrected for the distance and
foreground and host extinction; see below) of SN~2012ec, on the
plateau, is $M_{r^{\prime}}=-17.0\pm0.1$ mag; which is slightly brighter than normal Type IIP SNe in terms of $\sim R$-band
brightness \citep{arciip}, but it is clearly brighter than the
sub-luminous SN~2005cs \citep{pasto2005cs}.\\
Spectroscopic observations were conducted with a number of telescopes
and instruments: at $1$ and $2\mathrm{d}$ (post-discovery) using the ANU 2.3m telescope and the WiFeS integral field
spectrograph \citep{dopita07}, using the B3000/R3000 grism \citep[previously presented
  by][]{2012ec:spec}; at $3\mathrm{d}$ using the Nordic Optical
Telescope (NOT) using ALFOSC and the Gr\#4 grism; at $6\mathrm{d}$
with the Asiago 1.8m telescope using AFOSC and the Gr\#4; at $7\mathrm{d}$ and $15\mathrm{d}$ with the New
Technology Telescope using EFOSC and grisms Gr\#11 and Gr\#16 for the
first epoch and the Gr\#13 grism for the second epoch; and at
$9\mathrm{d}$ with the William Herschel Telescope using ISIS and
R300B and R158R grisms.  The sequence of spectroscopic observations is
shown on Figure \ref{fig:evo}.   The spectra presented in this paper are available in electronic format on WISeREP (the Weizmann interactive supernova data repository; \citealt{wiserep}\footnote{http://www.weizmann.ac.il/astrophysics/wiserep}).  At early times the spectrum of
SN~2012ec is similar to spectra of Type IIP SNe, being
dominated by broad P Cygni profiles associated with the hydrogen
Balmer series.  Fits conducted using GELATO \citep{gelato}\footnote{https://gelato.tng.iac.es/} identified similarities between the first spectrum of SN~2012ec and the spectrum of SN~1999em at $\sim 7$ days post-explosion \citep{2003MNRAS.338..939E} and earlier for other examples of Type IIP SNe, suggesting SN~2012ec was discovered $<6\mathrm{d}$ post-explosion.
At the first epoch, the velocity at the
$\mathrm{H\alpha}$ absorption minimum was measured to be
$-11\,700$\kms.  Over the sequence of spectra, the velocity of
\halpha\, decreased to $-9\,200$\kms by $15\mathrm{d}$ post-discovery.  The \halpha\, velocity, and its evolution, is similar to that
observed for SN~1999em \citep{2001ApJ...558..615H} and SN~2006bp
\citep[between 6 and 8 days post-explosion;][]{2007ApJ...666.1093Q}.  The velocities are significantly faster (by $\sim 4300\,{\mathrm{km\,s^{-1}}}$), and the rate of decline is slower, than observed for SN~2005cs at early times \citep{andrea05cs}.\\  
The unresolved Na I D doublet is observed in absorption at the
recessional velocity of NGC~1084 at all epochs.
We measure an average equivalent width $W_{\mathrm
  {Na\,I\,D}}=0.72\pm0.3$\AA.  Assuming Galactic-type dust, this
equivalent width corresponds to $E(B-V)=0.11$ mag
\citep{2003fthp.conf..200T} or $E(B-V)=0.10^{+0.15}_{-0.02}$ mag
\citep{2012arXiv1206.6107P}, and we adopt the latter for the total
reddening towards SN~2012ec (for identical Galactic and host
reddening laws; \citealt{ccm89}).
\section{Progenitor Observations and Analysis}
\subsection{Observations}
\label{sec:obs}
The pre- and post-explosion high spatial resolution observations of
the site of SN~2012ec are listed in Table \ref{tab:obs}.  Adaptive
optics observations were acquired with the VLT NACO instrument
on 2012 Aug 14.  The
observations were composed of dithered on-target exposures,
interleaved with offset sky images to aid with the removal of the sky
background, with the $K_{S}$-band filter.  The data were reduced using the {\it eclipse}
package\footnote{http://www.eso.org/sci/software/eclipse}.  The
observations used the S54 camera (with pixel scale
$0.0543\arcsec$) and, under the observing conditions, a FWHM of
$0.13\arcsec$ and Strehl ratio of $\sim 18\%$ were achieved.\\
The pre-explosion Wide Field Planetary Camera 2 ({\it WFPC2}) data
were retrieved from the STScI
archive\footnote{http://archive.stsci.edu/hst}, and were processed
with the most up-to-date calibration frames by the
On-the-fly-re-calibration (OTFR) pipeline.  The data were further
processed, combined and
photometry was conducted using the {\sc DOLphot}
package\footnote{http://americano.dolphinsim.com/dolphot/ v2.0} (with the $WFPC2$ module).  A further correction of $-0.1$
magnitudes was applied to the {\it WFPC2} photometry, to correct the
photometry to an infinite aperture \citep{1995chst.conf..269W}.\\
The Advanced Camera for Surveys Wide Field Channel ({\it ACS/WFC})
observations were also retrieved from the STScI HST data archive and
processed through the OTFR pipeline. The {\it ACS} observations was
composed of four separate dithered exposures.  We considered two
separate data products: the individual {\it FLT} distorted frames and the combined {\it DRC} frame, corrected for distortion and Charge Transfer
Inefficiency; photometry of these images was conducted using {\sc DOLphot} (using the $ACS$ module and the generic photometry mode, respectively).  The  distortion corrected images have a pixel scale of $0.05\arcsec$.  The derived
photometry was corrected to an infinite
aperture using the tabulated corrections of \citet{acscoltran}.

\begin{table}
\caption{\label{tab:obs}High-Spatial Resolution Observations of the Site of SN 2012ec}
\begin{tabular}{lrrr}
\hline\hline
Date         & Instrument          & Filter      & Exposure    \\
             &                     &             &  Time(s)    \\
\hline
2001 Nov 6   &  {\it HST/WFPC2/PC} & {\it F450W} & 320         \\
2001 Nov 6   &  {\it HST/WFPC2/PC} & {\it F606W} & 320         \\
2001 Nov 6   &  {\it HST/WFPC2/PC} & {\it F814W} & 320         \\
\\
2010 Sep 20  &  {\it HST/ACS/WFC}  & {\it F814W} & 1000        \\
\\
2012 Aug 14  &  VLT/NACO           & $K_{s}$      & 780 \\
\hline\hline
\end{tabular}
\end{table}

\subsection{Results}
\label{sec:res}

\begin{figure*}
\includegraphics[width=16.5cm]{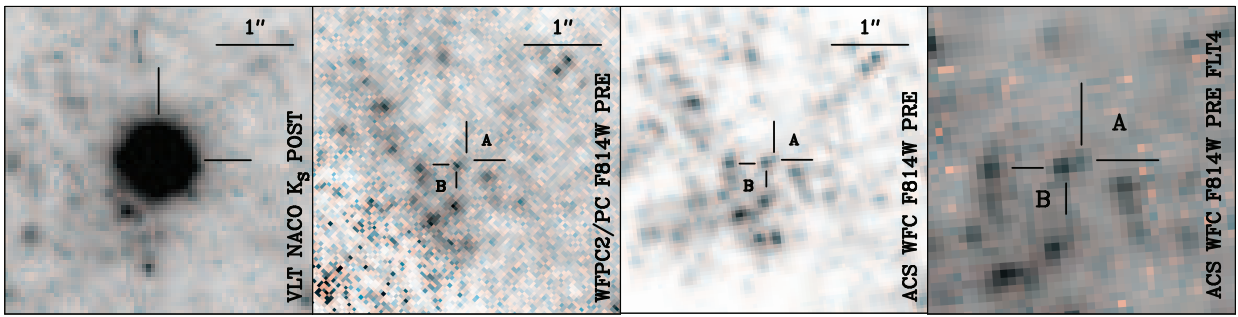}
\caption{High spatial resolution observations of the site of SN~2012ec (all images are oriented such that North is up and East is to the left).  {\it From l to r}: $4\arcsec \times 4\arcsec$ VLT NACO post-explosion image stamp centred on SN~2012ec; pre-explosion {\it HST/WFPC2/PC F814W} image, with the positions of sources A and B marked by cross-hairs; pre-explosion {\it HST/ACS WFC F814W} image; and a $\approx2\arcsec \times 2\arcsec$ section of a {\it distorted} pre-explosion {\it HST/ACS WFC F814W FLT} image (dataset jb4u02f0q\_flt - note the pixels are not square on the sky) centred on Source A.}
\label{fig:panel}
\end{figure*}

\begin{figure*}
\includegraphics[angle=0, width=16.5cm]{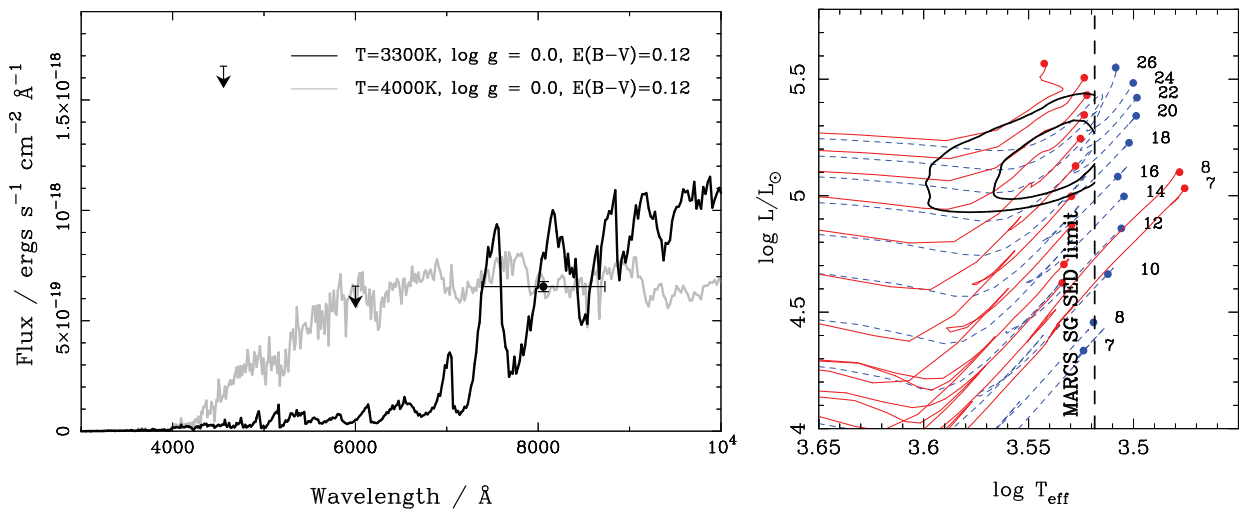}
\caption{The photometric properties of the progenitor candidate for
  SN~2012ec. {\it Left Panel)} The observed SED of the progenitor candidate Source A (composed
  of detections and limits in the {\it WFPC2} {\it F450W} and {\it
    F606W} and the {\it ACS/WFC} {\it F814W} bands) with MARCS spectra overlaid.  The upper limits represent
  the $99\%$-completeness limits derived from artificial star tests.
  {\it Right Panel)} The corresponding location of the progenitor
  candidate on the Hertzsprung-Russell Diagram, the contours
  correspond to 68\% and 95\% confidence intervals.  Overlaid are STARS
  stellar evolution tracks for solar (red) and twice solar (blue)
  metallicities.  The vertical dashed line corresponds the lowest
  temperature of the grid of MARCS spectra.}
\label{fig:sed}
\end{figure*}

Pre- and post-explosion images of the site of SN~2012ec are shown on
Figure \ref{fig:panel}.  The geometric transformation between the
post-explosion VLT/NACO image and pre-explosion {\it WFPC2} images was
calculated using 26 common stars, with a resulting r.m.s. uncertainty
on the transformation of $0.017\arcsec$. The transformed position of
the SN is located on the PC chip. Two sources are recovered in the
pre-explosion {\it WFPC2} images in close proximity to the transformed
SN position.  Source A, the progenitor candidate, is detected in the
{\it WFPC2 F814W} image only, offset from the transformed SN position
by $0.030\arcsec$ with brightness $m_{F814W}=23.39\pm0.18$ (detected
at $S/N = 5.9$).  The uncertainty on the astrometry for an object of
the faintness of Source A is $\sim 0.020\arcsec$ \citep{dolphhstphot},
such that the apparent discrepancy between the observed position of
Source A and the transformed SN position is not significant.  Source A is consistent with a point source with a measured sharpness value of -0.035 (good point-like sources should have sharpness values in the range -0.3 to 0.3).  Source B
is detected in the {\it F606W} and {\it F814W} images, offset from the
SN position by $0.11\arcsec$, with $m_{F606W}=23.00\pm0.06$ and
$m_{F814W}=22.23\pm0.08$ mags.  {\sc DOLphot} claims a detection
of Source B in the {\it F450W} image at $m_{F450W}=24.15\pm0.18$,
but on visual inspection of the image we suggest this detection is
questionable.  Source A is absent from the pre-explosion {\it WFPC2
  F606W} image; however, there is an extended source that overlaps
with the SN position in the pre-explosion {\it WFPC2 F450W} image.  It
is unclear if any of the flux observed in the {\it F450W} image arises
from Source A; given the scale of this feature, it is
unlikely that all of the flux is due to Source A alone.  Artificial
star tests were used to derive the detection limit at the SN position
in the pre-explosion {\it WFPC2} {\it F450W} and {\it F606W} images,
under the assumption that Source A does not contribute any flux in
these images.  We derived detection limits, corresponding to 50\%
recovery efficiency at $3\sigma$, of $24.5\pm0.3$ and $25\pm0.5$ mags in
the {\it F450W} and {\it F606W} filters respectively.\\
The geometric transformation between the post-explosion VLT/NACO image
and the pre-explosion {\it ACS/WFC F814W} was derived using 25 common
stars, with a resulting r.m.s. uncertainty of $0.016\arcsec$.  There
is clear evidence for flux, above the background, at the transformed
SN position, although {\sc DOLphot} photometry of the distortion
corrected {\it DRC} image identifies the flux as being extended
emission arising from Source B.
Inspection of all four individual {\it FLT} images, however, revealed that
this extended source is composed of two separate point sources,
coincident with Sources A and B (the offset between the transformed SN
position and Source A is $0.009\arcsec$).  We conclude
that Source B is not an extended source and that the object
observed in the {\it ACS/WFC} {\it DRC} images is a blend, arising from the {\it multidrizzle} image combination
process.  We
measured the {\it F814W} magnitudes for the progenitor candidate (Source
A) and Source B to be $23.10\pm0.04$ and $22.32\pm0.02$, respectively,
averaged over the independent photometry of the four {\it FLT}
images. Source A appears point-like in all four {\it FLT} images, with sharpness values of -0.072, -0.263, -0.075 and -0.064.\\
The brightness of Source B in the {\it ACS/WFC F814W} image is
approximately consistent with the magnitude of Source B measured in pre-explosion
{\it WFPC2 F814W} image.  There is a discrepancy between the
photometry of Source A in the {\it ACS} and {\it WFPC2} {\it F814W}
images, perhaps due to difficulties in partitioning the flux between
Sources A and B in the subsampled {\it WFPC2} image (due to the
relative faintness of Source A) and also the slight difference between the
{\it F814W} filters used by the two instruments. Using synthetic
photometry of MARCS SEDs \citep{marcsref}, we determined that, for
the coolest red supergiants, the colour between the {\it
  WFPC2} and {\it ACS} {\it F814W} filters is $\sim 0.1\,{\mathrm
  {mags}}$.\\
The observed photometry and upper limits
for the progenitor candidate (Source A) were compared with synthetic photometry of MARCS SEDs, using a Markov Chain Monte Carlo Bayesian inference scheme (Maund, 2012, in prep.).  We considered the $15M_{\odot}$ spherical MARCS models, with the effective gravity fixed at $\log g = 0.0$ \citep{2005ApJ...628..973L},
$\log Z/Z_{\odot}=+0.25$, and the reddening adopted in Section
\ref{sec:evo}, assuming only foreground and host extinction following a Galactic-type $R_{V}=3.1$ \citet{ccm89} reddening law.  
The corresponding allowed region on the Hertzsprung-Russell Diagram for the progenitor is shown on Figure \ref{fig:sed}.  The colour
constraint provided by the pre-explosion $F606W$ detection limit forces the resulting solution for the
progenitor to cool temperatures consistent with a red supergiant ($<4000K$).  The grid of
comparison MARCS SEDs, however, only goes down to $3300K$ and lower
temperatures, consistent with the end points of the twice solar
metallicity stellar evolution tracks, cannot be probed. We determine a
corresponding luminosity for the progenitor of
$\log(L/L_{\odot})=5.15\pm0.19$.  In comparison with the end points of
solar and twice solar metallicity STARS stellar evolution tracks
\citep{eld04}, we infer an initial mass for the progenitor of
$14-22_{\odot}$.  If the progenitor is subject to additional reddening, by local dust destroyed in the SN, we can only constrain a lower initial mass limit of $14M_{\odot}$.
In extrapolating the luminosity from the presented
contours to lower temperatures, we expect our derived mass estimate
would not change significantly.  The derived luminosity yields a
radius for the progenitor of $R=1030\pm180R_{\odot}$.

\section{Discussion and Conclusions}
\label{sec:conc}
We have presented the identification of the progenitor of a Type IIP SN, with an
initial mass in the range $14-22M_{\odot}$.  We have assumed that the extinction towards the progenitor is identical to the extinction affecting the SN.   \citet{2012MNRAS.419.2054W} and \citet{2012ApJ...759...20K}  explored the effect of dust formed in the winds of Red Supergiants, giving rise to additional extinction towards the observed progenitors of Type IIP SNe that is not probed by observations of surrounding stars or the SNe themselves (as this dust is destroyed in the subsequent explosion).   \citet{2012ApJ...759...20K} showed that this dust may also have a significantly different composition and, hence, reddening law to the dust of the interstellar medium \citep{ccm89}.  In the case of the
progenitor of SN~2012aw, which was detected at four different
wavelengths \citep{2012arXiv1204.1523F, van2012aw}, the constraints on the SED
were insufficient to break the degeneracy between effective
temperature and reddening.  
\citet{2012arXiv1204.1523F} found that the minimum reddening towards the
progenitor of SN~2012aw was significantly larger than the reddening
inferred towards the subsequent SN, suggesting circumstellar dust
around the progenitor was destroyed in the explosion.  In the likely event that the progenitor of SN~2012ec suffers from greater extinction than inferred from post-explosion observations of the SN, our inferred mass range translates to a lower mass limit for the progenitor of $>14M_{\odot}$.

The lower end of the inferred mass range  for the progenitor places it close to the maximum
threshold observed for the red supergiant progenitors of Type IIP SNe
\citep{2008arXiv0809.0403S}.  Late-time observations of the site of
SN~2012ec will be crucial to confirm the progenitor identification
(through its disappearance) and its derived mass; and will be crucial to
determining the nature of the extended emission observed in
the pre-explosion {\it WFPC2 F450W} image \citep{2009Sci...324..486M}.\\
Based on the early photometric and spectroscopic evolution of
SN~2012ec, we conclude that it is a slightly brighter than average Type IIP SN.
The similarities between the light curve and spectroscopic evolution
of SN~2012ec, with respect to early observations of other Type IIP
SNe, suggests it was discovered only a few days ($<6\mathrm{d}$)
post-explosion \citep{2012ec:spec}.  Unlike the probable yellow
supergiant progenitor of the Type II SN~2009kr
($M_{ZAMS}=15^{+5}_{-4}M_{\odot}$ - \citealt{2010ApJ...714L.280F};
$18-24M_{\odot}$ - \citealt{2010ApJ...714L.254E}), the apparent high mass of the progenitor of
SN~2012ec indicates that stars at the maximum mass threshold ($\sim
16M_{\odot}$) may explode as Red Supergiants and produce Type IIP rather than IIL SNe \citep{arciip}.
\section*{Acknowledgments}
Research by JRM is supported through a Royal Society Research Fellowship. SJS is supported by  FP7/2007-2013/ERC Grant agreement  n$^{\rm o}$ [291222].  GP acknowledges support from the Millennium
Center for Supernova Science through grant P10-064-F funded by
``Programa Bicentenario de Ciencia y Tecnolog\'ia de CONICYT'' and
``Programa Iniciativa Cient\'ifica Milenio de MIDEPLAN''. Research by AGY and his group is supported by funding from the FP7/ERC, Minerva and GIF grants. 
Based on
observations collected at the European Organisation for Astronomical
Research in the Southern Hemisphere, Chile and as part of PESSTO (the
Public ESO Spectroscopic Survey for Transient Objects Survey). The Liverpool Telescope is
operated on the island of La Palma by Liverpool John Moores University
in the Spanish Observatorio del Roque de los Muchachos of the
Instituto de Astrofisica de Canarias with financial support from the
UK Science and Technology Facilities Council (Program ID OL12B38).
\bibliographystyle{apj}

\end{document}